\documentclass{mn2e}

\usepackage{graphicx}
\usepackage{epstopdf}
\usepackage{times}

\newcommand{\nn}	{\mbox{N$_2$}}
\newcommand{\nf}	{\mbox{$^{15}$N}}

\newcommand{\nnhp}	{\mbox{N$_2$H$^+$}}

\newcommand{\delnf}     {\mbox{$\delta^{15}$N}}
\newcommand{\nfo}       {\mbox{$^{14}$N}}
\newcommand{\permil}    {\mbox{\fontencoding{U}\fontfamily{wasy}\selectfont\char104}}

\begin{document}

\title[Nitrogen superfractionation]{Nitrogen superfractionation in dense cloud cores}

\author[S. D. Rodgers and S. B. Charnley]
{S. D. Rodgers and S. B. Charnley \\
Space Science  \& Astrobiology Division, MS 245-3, NASA Ames Research
Center, Moffett Field, CA 94035, USA}

\maketitle

\begin{abstract}

  We report new calculations of interstellar $^{15}$N fractionation.
  Previously, we have shown that large enhancements of
  $^{15}$N/$^{14}$N can occur in cold, dense gas where CO is frozen
  out, but that the existence of an NH $+$ N channel in the
  dissociative recombination of N$_2$H$^+$ severely curtails the
  fractionation. In the light of recent experimental evidence that
  this channel is in fact negligible, we have reassessed the $^{15}$N
  chemistry in dense cloud cores.  We consider the effects of
  temperatures below 10~K, and of the presence of large amounts of
  atomic nitrogen. We also show how the temporal evolution of
  gas-phase isotope ratios is preserved as spatial heterogeneity in
  ammonia ice mantles, as monolayers deposited at different times have
  different isotopic compositions. We demonstrate that the upper
  layers of this ice may have $^{15}$N/$^{14}$N ratios an order of
  magnitude larger than the underlying elemental value. Converting our
  ratios to $\delta$-values, we obtain $\delta^{15}{\rm N} >
  3,000$~\permil\ in the uppermost layer, with values as high as
  10,000~\permil\ in some models. We suggest that this material is the
  precursor to the $^{15}$N `hotspots' recently discovered in
  meteorites and IDPs.
\end{abstract}


\section{Introduction}

Laboratory analyses of primitive solar system materials, such as
meteorites, interplanetary dust particles (IDPs), and cometary dust
particles returned by the {\it Stardust} mission, show anomalous
fractionation in the heavy isotopes of numerous elements
relative to that expected from the cosmic or solar system values
(Clayton \& Nittler 2004; Lodders \& Amari 2005; Alexander et al.\
2007; Mc\-Keegan et al.\ 2006). In the case of hydrogen and nitrogen,
the large D/H and \nf/\nfo\ ratios observed in some phases have been
attributed to the survival of D- and \nf-enriched material from the
interstellar medium (ISM; 
Alexander et al.\ 1998; Messenger 2000).
For deuterium, the observed ratios are consistent with models and
observations of the ISM, where low-temperature ion-molecule reactions
lead to enhanced D/H ratios in both gas- and solid-phase species
(e.g.\ Millar, Bennett \& Herbst 1989; Charnley, Tielens \& Rodgers
1997.)  However, there is little observational data on nitrogen
isotope ratios in the ISM, and models of the \nf\ fractionation in
typical dense clouds predict modest enhancements of $\sim 25$~per cent
(Terzieva \& Herbst 2000).  In comparison, the largest \nf\
enhancements detected in meteorites -- in so-called `hotspots' -- have
\nf/\nfo\ enhancements of more than a factor of four relative to the
Earth (i.e.\ $\delnf > 3000$~\permil \footnote{
$\delnf(\rm X) \equiv 1000 \times \left[ \left(\nf/\nfo\right)_{\rm X} / \left(\nf/\nfo\right)_\oplus
~ -1 \right]$ where $(\nf/\nfo)_\oplus$ is the terrestrial isotopic ratio of 0.003678 (De Bi\`evre et al.\ 1996)
 }; Busemann et al.\ 2006). Values of $\delnf > 1000$~\permil\  have also been
 found in hotspots in IDPs and {\it Stardust} samples
(Floss et al.\ 2004, 2006; McKeegan et al.\ 2006).

\begin{figure}
\begin{center}
\includegraphics[width=2.7in]{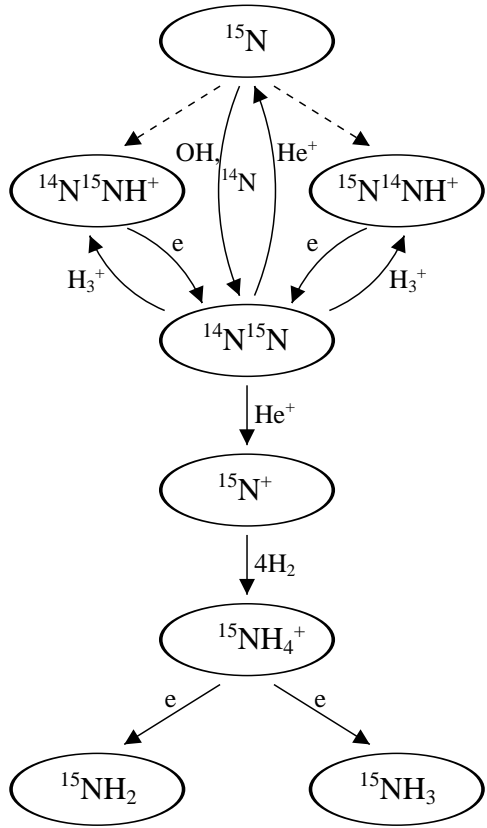}
\caption{Principal gas-phase reactions involving nitrogen. The dotted lines indicate the
fractionation reactions (\ref{frac1}) and (\ref{frac2}).}
\label{fig:bubble}
\end{center}
\end{figure}

In an earlier paper we demonstrated that significantly increased \nf\
fractionation can occur when CO is depleted onto dust grains (Charnley
\& Rodgers 2002 [Paper I]).
The key fractionation reactions are:
\begin{eqnarray}
{\rm ^{15}N + ~^{14}N_2H^+} ~&\rightleftharpoons&~ {\rm ^{14}N + ~^{15}N^{14}NH^+}
~~~ + \Delta E_1 \label{frac1}\\
{\rm \nf + ~^{14}N_2H^+} ~&\rightleftharpoons&~ {\rm ^{14}N + ~^{14}N^{15}NH^+}
~~~ + \Delta E_2 \label{frac2}
\end{eqnarray}
where the exothermicities are $\Delta E_1 = 27.7$~K and $\Delta E_2 =
36.1$~K (Terzieva \& Herbst 2000). These preferentially drive \nf\
into molecular nitrogen, at the expense of atomic N$^0$ which becomes
isotopically light.  In a standard dark cloud model
the degree of fractionation is limited by chemical reactions
which cycle nitrogen between atomic and molecular form (see Fig.~\ref{fig:bubble}).
%
However, if CO is frozen out as ice, OH is unavailable, this cycle is
broken, and much larger \nf-enhancements are possible. In this
scenario, degradation of the (\nf-enriched) N$_2$ by He$^+$ is
essentially a one-way process. The N$^0$ atoms released by this
reaction remain in the gas, whereas the N$^+$ ions react rapidly with
H$_2$ in a sequence of reactions to form ammonium
(Fig.~\ref{fig:bubble}).
The NH$_4^+$ ions then recombine with electrons, producing NH$_2$ and
NH$_3$, which subsequently freeze out on the dust grains.  If the
solid-phase NH$_2$ radicals are hydrogenated to NH$_3$, as expected
from models of grain surface chemistry (Brown \& Charnley 1990), the
end result is a large abundance of isotopically heavy ammonia ice. At
late times in this model, when ${\rm N^0/N_2} > 1$, the efficiency of
reactions (\ref{frac1}) and (\ref{frac2}) are greatly increased,
leading to much larger gas-phase \nf/\nfo\ ratios than in the standard
dense cloud chemistry.  Ultimately, we found that the ammonia ice is
enhanced in \nf\ by a factor of 1.8.

This model could account for the presence of high bulk $\delta \rm
^{15} N$ values in IDPs (Messenger 2000), and was also able to account
for the non-detection of \nn\ in comets.
Some additional processing was indicated to incorporate the $\rm ^{15}
N$-enriched ammonia into the carbonaceous matter, and secondary
energetic processing to add $\rm ^{15} NH_2$ side-groups to polycyclic
aromatic hydrocarbon (PAH) molecules was suggested. Recent meteoritic
studies have confirmed that most of the $\rm ^{15} N $ enrichment in
IDPs is carried by amine side-groups on aromatic moeties (Keller et
al.\ 2004).  However, the model was unable to reproduce the largest
\delnf\ values seen in the hotspots. Moreover, the experimental
determination by Geppert et al.\ (2004) that recombination of \nnhp\
preferentially breaks the N$\equiv$N bond, producing N$^0$ and NH,
acts to suppress the maximum fractionation (Rodgers \& Charnley 2004
[Paper II]).

In this paper, we are motivated to revisit our previous models as a
result of several recent discoveries.  Firstly, the experimental
results of Geppert et al.\ (2004) have been challenged by Molek et
al.\ (2007), who demonstrated that recombination of N$_2$H$^+$ leads
predominantly to $\nn + \rm H$, as indicated in previous work (Adams
et al.\ 1991). The new results indicate that rupture of the N$_2$ bond
occurs rarely, if at all, with an upper limit for the $\rm NH + N^0$
branching ratio of 5 per cent. Secondly, observations of several
pre-stellar cores have revealed that they have temperatures below 10~K
in their central, densest regions, as expected from physical models
(Evans et al.\ 2001; Zucconi, Walmsley \& Galli 2001). For example,
Crapsi et al.\ (2007) derived $T=5.5$~K in L1544, and Pagani et al.\
(2007) found $T=7$~K in L134N\@. Due to the small zero-point energy
changes associated with \nf\ fractionation, reactions (\ref{frac1})
and (\ref{frac2}) are extremely sensitive to temperature, and these
lower temperatures may be expected to yield larger \nf/\nfo\ ratios.

Thirdly, observations of \nnhp\ in dark clouds imply \nn\ abundances
of a few $\times 10^{-6}$ (Womack, Ziurys \& Wyckoff 1992; Maret,
Bergin \& Lada 2006). This is significantly less than the galactic
elemental nitrogen abundance and Maret et al.\ proposed that the
`missing' nitrogen was in atomic form. Although the chemical model of
Maret et al.\ assumed the Geppert et al.\ (2004) branching ratios for
N$_2$H$^+$ recombination, models assuming the `old' branching ratio
also predict ${\rm N}^0 > {\rm N}_2$ at early times (e.g.\ Pineau des
For\^ets, Roueff \& Flower 1990).  This is because $\rm N \rightarrow
N_2$ conversion occurs via neutral-neutral reactions, and the
time-scale for this process, $t_{\rm N_2}$, is larger than the
dynamical free-fall time, $t_{\rm ff}$\@. In contrast, CO is formed
fairly rapidly by ion-molecule chemistry, on a time-scale $t_{\rm
  CO}$, where $t_{\rm CO} < t_{\rm ff} < t_{\rm N_2}$.  Therefore, we
would expect young clouds to have high CO abundances but relatively small
${\rm N_2/N}$ ratios. This is consistent with the view that molecular
clouds and pre-stellar cores form rapidly, on a time-scale of order a
few times the free-fall time (Ballesteros-Paredes et al.\ 2007).
Dynamical-chemical models of collapsing clouds show that $\rm N
\rightarrow N_2$ conversion is never efficient over $t_{\rm ff}$
(Brown, Charnley \& Millar 1988). As discussed above, a large
N$^0$/\nn\ ratio is necessary to produce substantial
\nf-fractionation, so if the nitrogen in dense clouds is in fact
mainly atomic rather than molecular, this will have important
consequences for the \nf/\nfo\ ratios.

Finally, in our earlier work we only calculated the bulk isotope
ratios in the ammonia ice.  In reality, the grain mantles will have an
`onion-ring'-like structure consisting of sequentially accreted
monolayers. Thus, temporal variations in the gas-phase \nf/\nfo\
ratios will be preserved as spatial gradients in the ices, with each
layer recording the gas-phase ratio at the time it was accreted.  In
particular, the late-accreting, uppermost layers will be the most
highly fractionated.  As it is these layers which are likely to
experience the largest degree of subsequent processing, it is
necessary to distinguish between the bulk isotope ratios in the ice as
a whole, and those in specific monolayers.

\section{Model}

We utilize the same chemical model as in our previous work (Papers I
\& II). The \nf-fractionation chemistry is based on Terzieva \& Herbst
(2000); a list of the reactions and the rate coefficients we have
adopted appeared in Paper II\@. We consider a high density core
($n_{\rm H_2} = 5\times 10^6$~cm$^{-3}$) with temperatures of 5, 7,
and 10~K, and a cosmic ray ionization rate of
$5\times10^{-17}$~s$^{-1}$. The elemental \nf/\nfo\ ratio in each
model is generally assumed to be 1/400, although the effects of using
other values were also investigated.  A key parameter is the branching
ratio for the dissociative recombination of \nnhp:
\begin{eqnarray}
{\rm N_2H^+ ~+~ e} &~\longrightarrow~& {\rm N_2 + H} ~~~(1-f) \label{BR1}\\
&~\longrightarrow~& {\rm NH+ N} ~~~(f) \label{BR2}
\end{eqnarray}
We considered values for $f = 0.0$, 0.02, and 0.05, based on the
experimental results of Molek et al.\ (2007).  We use elemental
abundances for C, O, and N of 140, 290, and 80 parts per million
respectively, relative to hydrogen (Savage \& Sembach 1996), and
assume complete depletion of metals. All the carbon is initially in
the form of CO, and the remaining oxygen is atomic.

As before, we assume that CO freezes out onto grains with a sticking
coefficient of unity, whereas N and N$_2$ remain in the gas. The
reason for this selective depletion of CO versus N$_2$ is
controversial, as laboratory experiments reveal that the two species
have similar binding energies (\"Oberg et al.\ 2005). Nevertheless,
observations of pre-stellar cores show ample evidence for the presence
of N$_2$-rich, CO-poor regions toward the centers of these objects
(Bergin \& Tafalla 2007).  Due to their low polarizabilities, atoms
should have the lowest binding energies for physisorption to icy dust
grains. Reaction with an H atom forms a simple hydride that can stick
more effectively due to hydrogen bonding with the substrate. Thus, we
assume that O atoms are hydrogenated on the grains to form water
ice. Conversely, recent laboratory experiments indicate a very low
efficiency for the reaction of N$^0$ and H on ice surfaces (T Hiraoka,
private communication), which supports our assumption that N atoms do
not stick.

We have adapted our model to track the \nf/\nfo\ ratios as successive
monolayers (ML) of ammonia ice are accreted. A `typical' interstellar
grain of size 0.1~$\mu$m has $\sim 10^6$ surface sites, and an
abundance relative to hydrogen of $\sim 10^{-12}$. Hence, one ML
corresponds to a total solid phase abundance of $10^{-6}$, and our
elemental N abundance corresponds to a maximum of 80 ML of ice,
assuming that all the nitrogen freezes 
out as NH$_3$ (or NH$_2$)\@. In fact, a large fraction of the nitrogen
remains in the gas phase in the form of N$^0$, and we typically find
that we form $\sim 30$~ML of ammonia ice.

\section{Results}

We began by investigating the importance of the branching ratio, $f$, in
the recombination of \nnhp\@. We found only very small
differences in the results of models with $f$ equal to 5, 2, and 0 per cent,
and conclude that, as long as $f \leq 0.05$, the presence of channel 
(\ref{BR2}) does not significantly affect the \nf\ chemistry. We then varied the
elemental nitrogen isotope ratio, using values for \nfo/\nf\ of 800, 400, and 
100. In every case, the \nf\ enhancements are identical, relative to the
underlying ratio. Henceforth, we discuss models with $f = 0.02$ and $\nfo/\nf = 400$\@.

\subsection{Temperature dependence}
\label{sec:Temp}

\begin{figure}
\begin{center}
\includegraphics[width=3.2in]{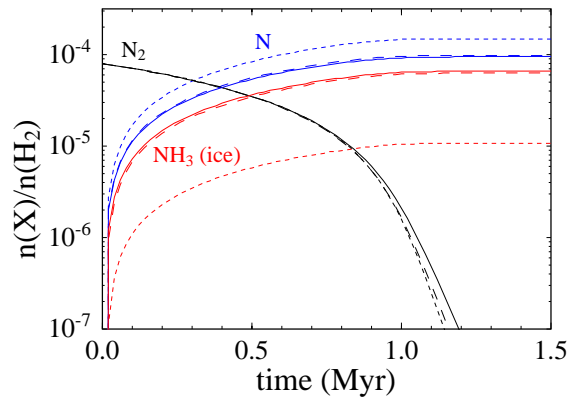}
\caption{Time evolution of the main nitrogen carriers when all the N
is initially molecular. Solid, dashed and dotted lines correspond to $T = 10$,
7, and 5~K respectively.}
\label{fig:NN2}
\end{center}
\end{figure}

Figure \ref{fig:NN2} shows the abundances of the three main repositories of nitrogen --
gaseous N$^0$ and N$_2$, and solid NH$_3$ -- as a function of time, assuming
that all of the nitrogen is initially present as N$_2$\@. The overall nitrogen chemistry
is the same at 10~K and 7~K: N$_2$ is degraded on a time-scale $\sim$~Myr.
At late times just over half of the total nitrogen is in the gas phase as atoms,
and just under half has frozen out as ammonia ice. At 5~K, however, very little
ice is formed, due to the endo-ergocity of the reaction
\begin{equation}
  {\rm N^+ + H_2} ~~\longrightarrow~~ {\rm NH^+ + H}
\label{eqn:n+h2}
\end{equation}
We have assumed an activation energy barrier of 85~K for this
reaction, based on the value in the UMIST reaction rate database
(Woodall et al.\ 2007).  At 5~K, the reaction becomes so slow that
radiative recombination becomes the dominant loss route for N$^+$
ions, and only small amounts of gas-phase NH$_2$ and NH$_3$ are
produced. In effect, the barrier for reaction (\ref{eqn:n+h2}) sets a
limit on the nitrogen chemistry in that, as the temperature drops, it
eventually becomes too cold to produce ammonia.  In extremely cold
cores the chemistry simply transforms the initial gas-phase N$_2$ into
gas-phase N$^0$\@. In the following, we therefore look at the
fractionation in 7~K and 10~K gas.

\begin{figure*}
\begin{center}
\includegraphics[width=6.8in]{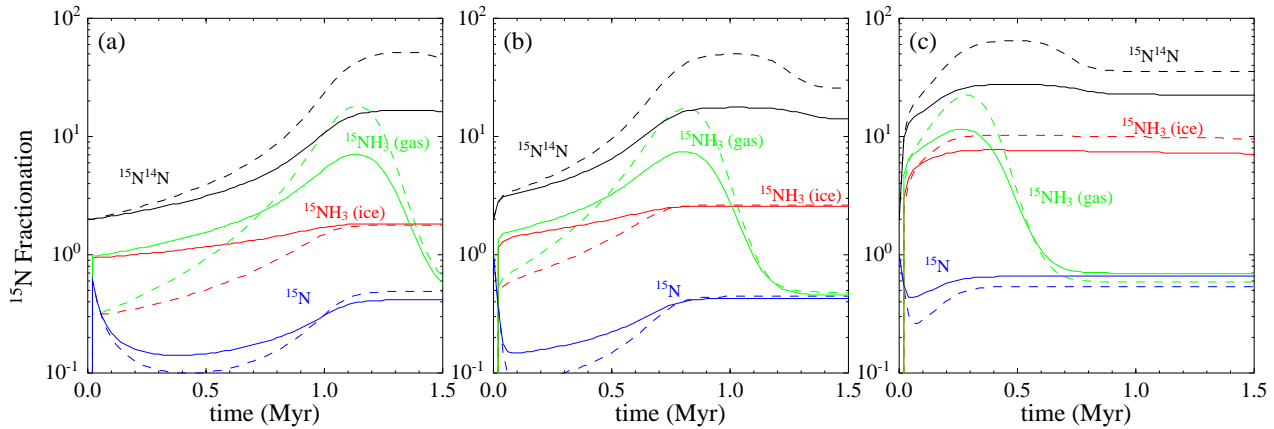}
\caption{Time evolution of the \nf/\nfo\ enhancement ratios in key
  species at $T= 10$~K (solid lines) and $T = 7$~K (dashed lines). The
  initial fraction of nitrogen in molecular form is 100\%, 63\%,
  and 10\% in panels (a), (b), and (c) respectively.}
\label{fig:tfrac}
\end{center}
\end{figure*}

Figure~\ref{fig:tfrac}a shows the fractionation in the most important
molecules.  The results for 10~K gas are similar to those shown in
Paper I, whereas at 7~K significantly larger peak \nf/\nfo\ ratios are
found in gas-phase NH$_3$ and N$_2$\@.  At early times, the gas-phase
ammonia formed at 7~K is actually {\it depleted} in \nf\@. This is due
to the reduced rate of reaction (\ref{eqn:n+h2}) which leads to a much
larger N$^+$ abundance than at 10~K\@. These N$^+$ ions undergo
isotope-exchange reactions with N$_2$ (Freysinger et al.\ 1994):
\begin{equation}
{\rm ^{15}N^+ + ~^{14}N_2} ~~~\rightleftharpoons~~~ {\rm ^{14}N^+ + ~^{15}N^{14}N}
~~~ + \Delta E_3 \label{frac3}
\end{equation}
where $\Delta E_3 = 28.3$~K (Terzieva \& Herbst 2000). This reaction
shuffles \nf\ back into N$_2$, reducing the \nf-fractionation in N$^+$
and thus ammonia. At late times, this effect is still suppressing the
N$^+$ fractionation relative to N$_2$, but the enormous
\nf-enhancements in N$_2$ ensures that N$^+$ and NH$_3$ are more
fractionated than at 10~K\@. In terms of the bulk \nf/\nfo\ ratio in
the ammonia ice, both the 10~K and 7~K models predict similar
enhancements of $\approx 1.8$\@. However, as discussed earlier, the
time evolution of the isotope ratios will be preserved in the layered
structure of the ices. Even though the overall icy \nf/\nfo\ ratios
are the same in both models, ammonia ices which accreted at 7~K will
have a more heterogeneous structure, with the innermost layers
depleted in \nf, but with the top layers enhanced by factors $\sim
10$.

\subsection{The initial N/N$_2$ ratio}

Significant fractionation in N$_2$ can only occur when N$_2$ is not
the dominant nitrogen carrier. In the models discussed so far, this
occurs at late times when most of the N$_2$ has been broken down by
He$^+$ attack, on time-scales $\sim {\rm Myr}$. However, if a
substantial fraction of the nitrogen is initially atomic, enhanced
\nf/\nfo\ ratios can be produced on much shorter time-scales. The
N$^0$/N$_2$ ratio at $t=0$ depends on the dynamical and chemical
history of the cloud before the formation of the dense pre-stellar
core, and can be taken as a free parameter in our model. To
investigate the importance of this factor, we have considered three
additional models, where the initial fraction of N in molecular form
is taken to be 0.6, 0.1, and 0. The latter case, where all of the
nitrogen is initially atomic, is not particularly realistic, but was
used to constrain the amount of N$_2$ that can be synthesized from
N$^0$ in the limited time available before CO etc.\ freeze out. We
find that this produces only very small N$_2$ abundances, $\approx 2$
per cent of the total nitrogen. This means that the available pool of
N$_2$ -- roughly half of which will end up as ammonia ice -- is
essentially equal to the N$_2$ abundance at $t = 0$.

Figure \ref{fig:tfrac}b and \ref{fig:tfrac}c shows the isotope
chemistry when a large fraction of the original nitrogen is
atomic. Qualitatively, the gas-phase \nn\ and NH$_3$ fractionation is
similar to that in fig.~\ref{fig:tfrac}a, except that the peak
\nf/\nfo\ ratios occur much earlier as the initial N$_2$ abundance is
reduced. As before, we find that the upper layers of the ice should be
enhanced in \nf\ by an order of magnitude.  In terms of the overall
bulk isotope ratio in the ice, it is clear that much larger values
result, with $\rm (^{15}NH_3/^{14}NH_3)_{ice} = 7$ when only ten per
cent of the initial nitrogen is molecular. This is due to the fact
that less ammonia ice is formed overall, so the highly-fractionated
upper layers account for a greater proportion of the total ice.  We
find roughly the same peak ratios regardless of the initial
N$^0$/N$_2$ ratio, but in models with more N$_2$ the bulk
fractionation is diluted by the large number of monolayers which
accrete at early times with essentially normal (or even reduced in the
7~K model) \nf/\nfo\ ratios.

\subsection{Isotope ratios in individual monolayers}

\begin{figure*}
\begin{center}
\includegraphics[width=6.8in]{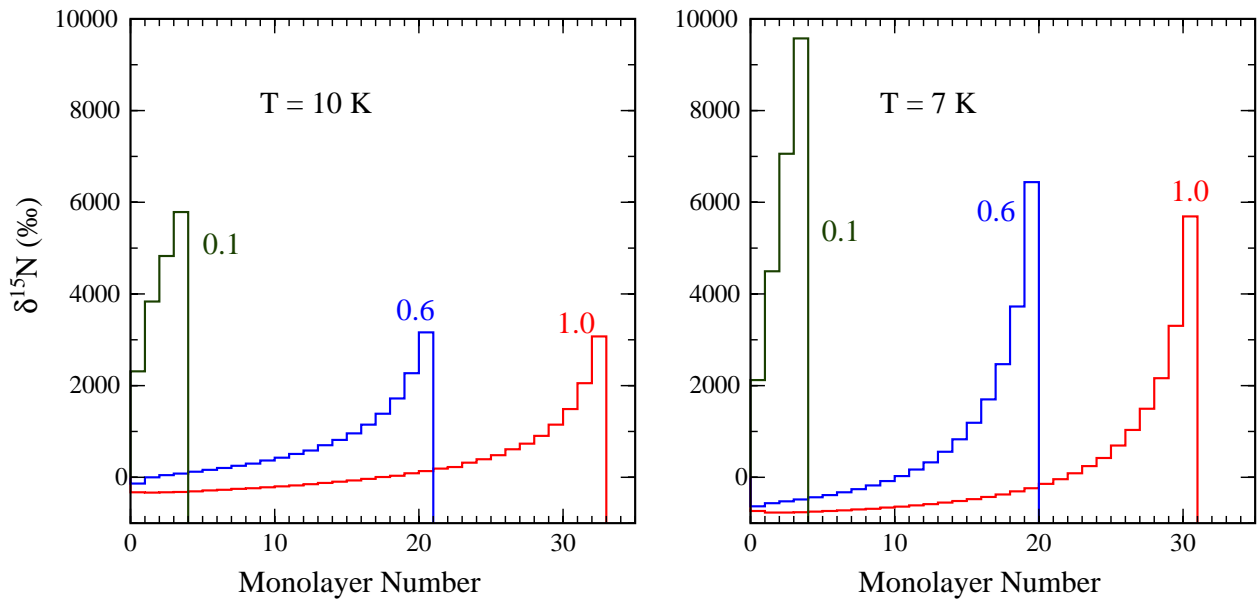}
\caption{$\delta$\nf\ values in different
monolayers, for $T = 10$~K and $T = 7$~K\@. Labels refer to the fraction of nitrogen initially assumed
to be molecular.}
\label{fig:ML}
\end{center}
\end{figure*}

We have calculated the \nf/\nfo\ in successive ML as they accrete from the gas.
For comparison with laboratory measurements, we have converted the
isotopic ratios into $\delta$-values relative to the terrestrial
\nf/\nfo\ ratio. In order to
do this, we need to know the original \nf/\nfo\ ratio in the material from which
the protosolar nebula (PSN) was formed. We assume a value of
$(\nf/\nfo)_{\rm PSN} = 0.0025$, which is implied
by three independent measurements: Jupiter's atmosphere, high-temperature nebular condensates,
and the solar wind (Fouchet et al.\ 2004; Meibom et al.\ 2007; Kallenbach, Bamert \& Hilchenbach 2007.)
Figure \ref{fig:ML} shows the $\delta$-values in different ML for
several models.
Clearly, the upper layers of the ice are the most highly fractionated, with peak
values of $\delta\nf > 3000$~\permil\ in every case. Also apparent is the approximate
proportionality between the initial N$_2$ abundance and the total number of ML
that accrete. When only ten per cent of the nitrogen is initially molecular, only four
ML of NH$_3$ ice are formed. However, in this case every ML is very highly fractionated,
with $\delta\nf$ everywhere greater than 2000~\permil. As discussed in $\S$\ref{sec:Temp},
the peak values are larger for $T = 7$~K, but the 7~K model also produces a large
number of ML at early times that are slightly depleted in \nf.

\section{Conclusions}
  
We have revisited our earlier work on \nf\ fractionation based on
recent experimental and observation results. Using the branching
ratios for N$_2$H$^+$ dissociative recombination derived by Molek et
al.\ (2007), we effectively recover the results of our earlier work in
dense, CO-depleted gas at 10~K (Paper I\@.) We have investigated the
effects of varying the temperature, and show that at lower
temperatures, larger \nf/\nfo\ ratios are produced in gas-phase in
N$_2$\@. However, the barrier for the reaction of N$^+$ ions with
H$_2$ sets a lower limit on the temperature at which ammonia can be
produced efficiently. Assuming the `standard' rate coefficient for
this reaction we find that very little ammonia ice is generated for $T
< 7$~K\@.  We have also looked at the effects of a substantial
N$^0$/N$_2$ ratio at $t = 0$.  We find that, because roughly half of
the initial N$_2$ ends up in the form of NH$_3$ and NH$_2$, reduced
molecular nitrogen abundances yield less ammonia ice in total. Smaller
N$^0$/N$_2$ ratios do not significantly affect the peak gas-phase
fractionation ratios, but because the highly-fractionated ammonia
formed at late times represents a greater proportion of the total ice,
we find that the bulk ice \nf/\nfo\ ratio can be greatly increased.

Following the \nf/\nfo\ ratios in individual monolayers as they
accrete sequentially from the gas, we have shown that gas-phase
temporal variations in isotopic ratios are preserved as spatial
gradients in the layered ammonia ice. The uppermost layers which
accrete at late times have the largest \nf-enhancements, up to an
order of magnitude with respect to the elemental \nf/\nfo\ ratio.
Converting to $\delta$-values to compare with laboratory
determinations of the isotope ratios in primitive solar system
materials, we derive peak values of $\delta ^{15}{\rm N} >
3000$~\permil\ in gas at 10~K, and values as large as $\delta
^{15}{\rm N} \sim 10000$~\permil\ in models with lower temperatures
and smaller N$^0$/N$_2$ ratios. This is more than sufficient to
account for the largest measured ratios, and demonstrates that
interstellar gas-phase chemistry is likely the ultimate source of
cometary and meteoritic \nf\ anomalies.


\vspace{4mm}

\noindent
This work was supported by NASA's Origins
of Solar Systems Program through NASA Ames cooperative agreement
NNX07AO86A with the SETI Institute, 
and by the NASA Goddard Center for Astrobiology.


\newcommand{\apj}{ApJ}
\newcommand{\mnras}{MNRAS}
\newcommand{\aap}{A\&A}


\end{document}